\documentclass[10pt,twocolumn,footinbib]{revtex4}
\usepackage{graphicx,verbatim}
\usepackage{bm}
\usepackage{hyperref}
\usepackage{color,graphics,graphics,epsfig,amsmath,amssymb}
\usepackage{subfigure}
\hypersetup{urlcolor=blue, colorlinks=true}  

\def\priorIhst{$\Omega_k = 0.002 \pm 0.009$}
\def\priorIInohst{$\Omega_k = 0.013 \pm 0.012$}
\def\priorInohst{$-0.12 < \Omega_k < 0.01$} 
\def\BASIC{{\sc BASIC~}}

\def\Hunit{{\rm ~km s^{-1} Mpc^{-1}}}
\def\fig{Fig.~}
\def\p{{\cal P}}

\newcommand{\etal}{{\it et al.}}

\begin{document}

\title{How Flat is Our Universe Really?}
\author{P. M. Okouma$^{1-4}$, Y. Fantaye$^{5,6}$ and B. A. Bassett$^{1-4}$  \\
\it $^1$ Department of Maths and Applied Maths, University of Cape Town, Rondebosch 7701, Cape Town, South Africa \\
\it $^2$ South African Astronomical Observatory, Observatory, Cape Town, South Africa\\
\it $^3$ African Institute for Mathematical Sciences, 6-8 Melrose Road, Muizenberg, Cape Town, South Africa \\
\it $^4$ Centre for High Performance Computing, 15 Lower Hope St., Rosebank, Cape Town, South Africa\\
\it $^5$ Astrophysics Sector, International School for Advanced Studies, SISSA, 34136 Trieste, Italy\\
\it $^6$ Institute of Theoretical Astrophysics, University of Oslo, P.O.Box 1029 Blindern, N-0315, Oslo, Norway}

\begin{abstract}
Distance measurement provide no constraints on curvature independent of assumptions about
the dark energy, raising the question, how flat is our Universe if we make no such assumptions? Allowing for general evolution of the dark energy equation of state with 20 free parameters that are allowed to cross the phantom divide, $w(z) = - 1$,  we show that while it is indeed possible to match the first peak in the Cosmic Microwave Background with non-flat models and arbitrary Hubble constant, $H_0$, the full WMAP7 and supernova data alone imply \priorInohst ($2\sigma$). If we add an $H_0$ prior, this tightens significantly to \priorIhst. These constitute the most conservative and model-independent constraints on curvature available today, and illustrate that the curvature-dynamics degeneracy is broken by current data, with a key role played by the Integrated Sachs Wolfe effect rather than the distance to the surface of last scattering. If one imposes a quintessence prior on the dark energy ($-1 \leq w(z) \leq 1$) then just the WMAP7 and supernova data alone force the Universe to near flatness: \priorIInohst. Finally, allowing for curvature, we find that all datasets are consistent with a Harrison-Zel'dovich spectral index, $n_s = 1$, at $2\sigma$, illustrating the interplay between early and late-universe constraints. 
\end{abstract}

\maketitle

%\vspace{-0.2in}

{\it Introduction --} The sign and magnitude of the cosmic curvature, $\Omega_k$, is one of the most fundamental characteristics of our cosmos. The sign controls the default topology of the universe while the magnitude has real importance in testing theories: eternal inflation would be seriously tested if $|\Omega_k| > 10^{-4}$ \cite{Guth12,Kleban12} while anthropic considerations suggest that $|\Omega_k|$ might be large enough to be detectable \cite{Freivogel06}. Assuming $\Lambda$CDM is correct, we are not far from reaching this interesting regime with the latest curvature constraints around the $\sigma_{\Omega_k} \simeq 10^{-3}$ level \cite{Sanchez2012,lado2012}. Although there is significant prior dependence in these constraints \cite{Vardanyan_Trotta2011} as we will discuss in detail below, they represent a huge improvement over the major break-through from the BOOMERANG mission a decade ago which gave $|\Omega_k| \leq 0.2$ \cite{Boomerang00}, itself an order of magnitude improvement over earlier constraints \cite{Tegmark1999}.

The key step forward in measuring curvature was the realisation that the position of the first acoustic peak in the Cosmic Microwave Background (CMB) provides a standard ruler, and hence the angular diameter distance to the surface of last scattering \cite{Hu1995}. However, this alone does not constrain $\Omega_k$ because of the well-known geometric degeneracy between $\Omega_k$ and the Hubble parameter today, $H_0$ \cite{geomdeg}. With the addition of current large scale structure data, however, this degeneracy is now almost completely broken. 

In this paper we are concerned, however, with the more pernicious degeneracy between $\Omega_k$ and dark energy dynamics, parametrised through the equation of state, $w(z)$, which even an infinite number of perfect distance measurements cannot break \cite{weinberg70, Bernstein05, wright, CCB, HCCB, Mortsell11}. It follows from simple degrees of freedom counting: we are attempting to measure a free function - the expansion rate, $H(z)$ - as well as a constant, $\Omega_k$, from a single free function, the distance $d_A(z)$. Such degeneracies are peculiar neither to distances nor curvature alone \cite{HCCB}, as illustrated by the dark matter-dark energy degeneracy which makes it impossible to measure $\Omega_m$ without assumptions about dark energy\cite{kunz}. Nevertheless, the curvature of our universe is a geometric characteristic independent of the gravitational dynamics of spacetime and hence is particularly fundamental. 

\begin{figure}[h!]
\includegraphics[width=3.8in]{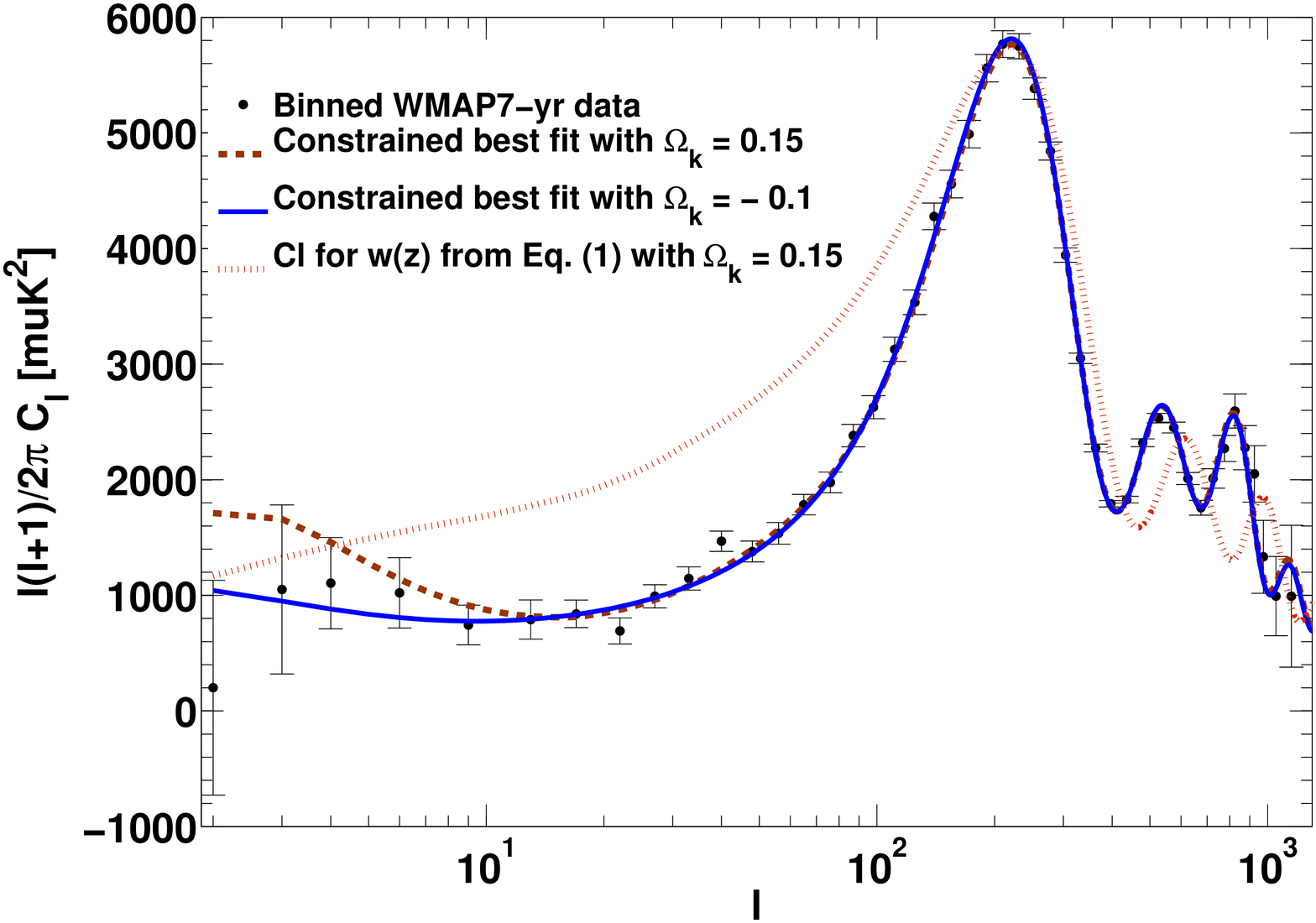}
\caption{Temperature angular power spectra for various curved models along with the binned WMAP7 data. The blue solid curve shows $C_{\ell}$ for the best-fitting model with $\Omega_k = -0.1$. The brown dashed curve shows the best-fit from an MCMC search with fixed $\Omega_k = 0.15$. It has $H_0 = 56.4\Hunit$. The fit is essentially perfect for $\ell > 10$ but the overall fit is poor due to the large Integrated Sachs Wolfe effect which disfavors all significantly open models. The red dotted curve corresponds to a model with $\Omega_k = 0.15$ and $H_0 = 71\Hunit$ and uses the $w(z)$ from Eq. (\ref{wz}) that  matches the flat $\Lambda$CDM distances at all redshifts. Although the first peak matches perfectly, the remaining fit is very poor. These examples show that the distance to the surface of last scattering is not important in our constraints on curvature.}
\label{cls}
%\vspace{-0.15in}
\end{figure}

The dark energy equation of state, $w(z)$, which will make a universe with curvature parameter $\Omega_k$ exactly mimic the distances in a flat, $\Lambda$CDM model at all redshifts, is given by \cite{CCB}: 
 \begin{widetext}
 \begin{eqnarray}
 \tiny
 w(z) = \frac{2}{3} \left[  (1+z)\left\{[\Omega_{ k} {D}_{ L}^2+(1+z)^2]{D}_{
 L}''-\frac{1}{2}(\Omega_{ k} {D}_{
 L}'^2+1)[(1+z){D}_{ L}'-{D}_{ L}]\right\} \right ] \nonumber \\
 \left[  [(1+z){D}_{
 L}'-{D}_{ L}]\left\{(1+z)[\Omega_{ m}(1+z)+\Omega_k]{D}_{
 L}'^2-2[\Omega_{ m} (1+z)+\Omega_{ k}] {D}_{ L}{D}_{
 L}'+\Omega_{ m} {D}_{ L}^2-(1+z)\right\} \right]^{-1}   \, , 
 \label{wz}
 \end{eqnarray}
 \end{widetext}
where $ {D}_L=(H_0/c) d_L $ is the dimensionless luminosity distance in the flat $\Lambda$CDM model and primes denote redshift derivatives. Figure (\ref{cls}) shows this degeneracy via the dotted red curve CMB power spectrum for an open model with $\Omega_k = 0.15$ with $w(z)$ given by Eq. (\ref{wz}) showing that it matches the position of the first peak, and hence the distance to the surface of last scattering, perfectly, although of course it is a bad fit to the rest of the data. 

Only by making assumptions about the dark energy dynamics can any distance measurements, even if measured perfectly at all redshifts, provide constraints on curvature. However, such constraints are fictitious unless we know the true nature of dark energy, which is not the case today. The strongest constraints arise, of course, by assuming that the dark energy is a cosmological constant, $\Lambda$ with $w(z)=-1$. 

\begin{figure}%[h!]
\centering
\includegraphics[width=3.8in]{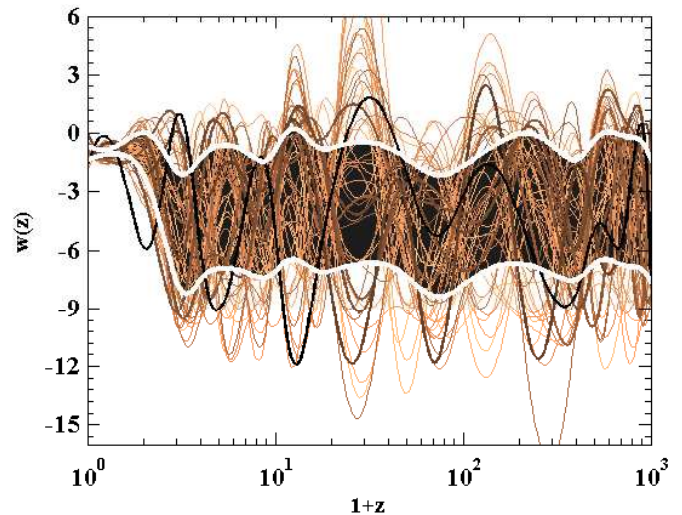}
\caption{ 50 randomly-selected dark energy $w(z)$ curves from the converged chains with Prior I ($-15 \leq w_i \leq 1$) and the \BASIC (WMAP7+SNIa+BBN) + HST data package. While $w(z)$ is well-constrained at low redshift, it is unconstrained at $z > 1$. Darker, thicker curves have higher likelihood. The dark band bounded by the white curves shows the splined 1-$\sigma$ confidence region for $w(z)$. At $2\sigma$ essentially the entire prior volume is available for $z > 1$.}
\label{w_cls_blocs}
\end{figure}

Since essentially all current constraints on curvature derive from distance measurements from the CMB, supernovae and Baryon Acoustic Oscillations (BAO) with restrictive assumptions about the dark energy, a natural question is `how curved is our Universe allowed to be if we do not assume anything about the dark energy dynamics'? Could we fit the data with $\Omega_k = 1.5$ with a suitable $w(z)$ for example? 

There are two relevant issues to answering this. The first is the freedom in the $w(z)$ parametrisation: how many free $w_i$ parameters? The second is the range over which the $w_i$ parameters are allowed to vary. To match flat $\Lambda$CDM distances with a closed model requires a $w(z)$ that typically must go out of the range $-1 \leq w(z) \leq 1$ allowed by the weak energy condition and naive non-superluminal speed of sound \cite{CCB}.  The $w(z)$ freedom dictates how subdominant dark energy was to matter and radiation when these components dominate the universe. Without enough $w(z)$ freedom a distance measurement during matter domination in addition to the CMB does break the curvature degeneracy \cite{Knox06}. 
 
{\it Method --} We  used the  CosmoMC Markov Chain  Monte Carlo (MCMC) package  \cite{Lewis} together with the WMAP7 likelihood for our parameter estimation and typically used five chains of about $3 \times 10^5$ steps each and used the Gelman-Rubin test for convergence.  To allow for general dark energy evolution, we used the  PPF  module  \cite{ppf} which splines an arbitrary number of parameters, $w_i$, at arbitrary redshifts to give $w(z)$ and crucially, can deal properly with perturbation evolution across the phantom divide, $w(z) = -1$. Our results are not very sensitive to the total number of $w_i$ parameters. This is not surprising since we effectively have only a few accurate distance measurements at $z < 2$ and one at $z \simeq 1100$; compatible with Fisher matrix projections \cite{MUH}. We simply need to ensure that we have enough degrees of freedom for  $w(z)$ to capture the degeneracy. All results in this paper are quoted for 20 $w_i$ spline coefficients located at approximately logarithmically-spaced redshifts allowing for rapid $w(z)$ variation at low redshift. 
% : http://arxiv.org/abs/1004.0236,0810.1744v1 : MUH
We considered two classes of dark energy priors on the $w_i$. {\bf Prior I} was flat over the range $w_i \in [-15,1]$ \footnote{We also ran chains with $w_i \in [-15,15]$ but were unable to achieve convergence due to the large available phase space but tiny volume of acceptable models with $w(z) > 1$.} while {\bf Prior II} was flat over $ [-1,1]$, mimicking quintessence-type models \cite{CCB}. We assume adiabatic initial conditions and used $k_0  = 0.002~{\rm Mpc^{-1}}$ as the pivot scale  for scalar and tensor power spectra
normalization \cite{Larson11}. The parameter set we used in our MCMC chains, which includes the 9 standard cosmological parameters in addition to our 20 $w_i$, is: $\p =  (\Omega_b h^2, \Omega_c h^2, [\Theta  \ or \  H_0], \tau,
\Omega_k,  n_s,  A_s,  r,  A_{sz}, {w_i})$. 

\begin{figure}
\begin{flushleft}
 \includegraphics[width=3.7in]{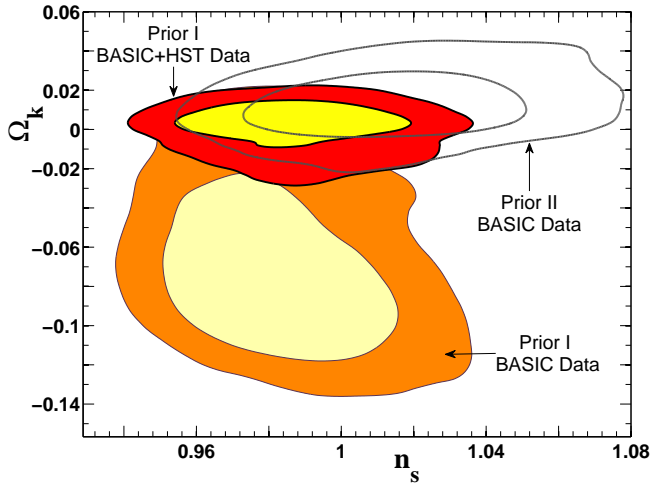}
\caption{Joint contours on $\Omega_k$ and $n_s$ for the three data sets and priors: \BASIC + Prior I ($-15 \leq w_i \leq 1$), \BASIC + {\sc HST} + Prior I, \BASIC + Prior II ($-1 \leq w_i \leq 1$). Here \BASIC = {WMAP7 + SNIa + BBN}. Even allowing for general dark energy models and conservative data choice constrains the curvature to satisfy \priorIhst, but all cases are compatible with $n_s = 1$. {One might suspect that the strong constraining power of  $H_0$ implies that the constraints on $\Omega_k$ are primarily coming from the supernovae data at low redshifts. In fact $H_0$ is important primarily because it controls the Hubble scale which sets the overall scale for $d_L(z)$.}}
\label{okns}
\end{flushleft}
\vspace{-0.25in}
\end{figure}

We used two minimal sets of data that are chosen to be maximally independent of assumptions about curvature and dark energy. The \BASIC data package consists of the CMB temperature and polarization data from WMAP7 \cite{Komatsu11}, the Union2 supernovae \cite{Amanullah10} and the Big Bang  Nucleosynthesis (BBN) prior $\Omega_b h^2 = 0.020 \pm 0.002$ (2$\sigma$) \cite{Burles01}. We do not include small-scale CMB data since getting the distance to the surface of last scattering right ensures that the $\ell > 1000$ data are well-fit and hence provide no additional constraining power. 

The \BASIC + HST data package adds the Hubble Space Telescope constraint \cite{Riess09} on the Hubble constant, $H_0 = 74.2 \pm 3.7 \Hunit$ \footnote{Although there is a measurement of $H_0$ with smaller errorbar \cite{Riess2011}, this is in marginal tension with recent BOSS results \cite{BOSS} so we prefer the broader constraint in order to be conservative.}. We do not include BAO and growth information to be conservative, since the assumption of flatness and $\Lambda$CDM comes in subtly, for example in the N-body simulations used to calibrate results. As we now show, we do not need extra data to rule out significant curvature even in the case of the most general $w(z)$.  
  
{\it Results} -- { In \fig(\ref{w_cls_blocs}) we show a random selection of 50 splined w(z) curves from our chains with BASIC + HST data. Here darker and thicker curves have higher likelihood. We also plot the splined 1-$\sigma$ limits (shown by the thick white curve bounding the shaded region) on $w(z)$ showing that current data give very little constraints on $w(z)$ for $z > 1$, essentially being determined by the prior.} 
Three relevant CMB spectra are shown in \fig(\ref{cls}). 
%The global best-fit with Prior I and the \BASIC data package which has $\Omega_k = -0.03$ is shown as the solid blue curve.
The global best-fit with Prior I and the \BASIC data package where $\Omega_k = -0.1$ is shown as the solid blue curve. %PO
The best fit with fixed $\Omega_k = 0.15$ and $H_0 = 71 \Hunit$ (brown dashed curve) shows why open models are ruled out by the Integrated Sachs Wolfe (ISW) effect at $\ell < 20$. This is not surprising since to match flat $\Lambda$CDM distances typically requires very rapid $w(z)$ evolution at low redshifts \cite{CCB} with resulting large ISW effect \cite{REES68,Crittenden96, psisw}. The $C_{\ell}$'s of the model with fixed $\Omega_k = 0.15$, WMAP7 best-fit parameters and the $w(z)$ corresponding to Eq. (\ref{wz}) which ensures that the distances are identical to the best-fitting flat $\Lambda$CDM model at all redshifts is shown by the red dotted line. Clearly the distance to the surface of last scattering plays little direct role in constraining curvature in general, unlike the case for $\Lambda$CDM.  
% {\bf Of all the low redshift data used, it has the strongest impact on curvature constraints. [P.O.]}
Our main finding is that with the \BASIC + HST data we recover a constraint on the curvature:  \priorIhst, even with effectively no limits on the $w_i$ (Prior I). The HST prior on the Hubble constant is critical in removing the closed branch of universes that are excellent fits to the \BASIC data package and have low $H_0$.  Indeed, without the HST prior the $\Omega_k$ posterior peaks around $\Omega_k = -0.085$. Assuming quintessence-like dark energy (Prior II) also removes the closed branch, irrespective of the HST constraint. The importance of allowing crossing of the phantom divide, $w(z) = -1$, is a consequence of trying to match distances, since Eq. (\ref{wz}) requires $w \rightarrow -\infty$ at some redshift when $\Omega_k < 0$. These results are exemplified in \fig(\ref{okns}) and summarised in Table (\ref{cosmic_deg}).   

\begin{table*}[htb]
%\vspace{-0.2in}
\scriptsize
\begin{center}
\begin{tabular}{lll llllll}
\hline
\hline
\textbf{Class} & \textbf{Parameter} &  \textbf{WMAP7} &  \textbf{WMAP7} & {\sc \bf BASIC}+
& {\sc \bf BASIC}+ & \textbf{{\sc \bf BASIC}+HST} & {\sc \bf BASIC+ACT} & \textbf{{\sc \bf BASIC}+ACT} \\
 & & \textbf{$\Lambda$CDM} & \textbf{O$\Lambda$CDM} & \textbf{Prior II}
& \textbf{Prior I} & \textbf{Prior I} & \textbf{+Prior I} & \textbf{+HST+Prior I} \\
\hline
\hline
Primary &  & \\ & $w(z=0)$  & & & $  -0.867 \pm 0.107 $  & $-1.043 \pm
0.538$ & $  -1.026 \pm 0.342 $ &  $ -0.999 \pm 0.554 $ &  $ -1.061 \pm
 0.349 $ \\
 & $\Omega_k$ & & $-0.080^{+0.071}_{-0.093}$ &  $ 0.013 \pm 0.012 $ & $
 -0.069^{+0.035}_{-0.033}  $ & $ 0.002 \pm 0.009  $ & $ -0.056 \pm  0.035 $ & $
 0.002  \pm  0.009  $  \\  &  $H_0$  [km/s/Mpc] &  $71  \pm  2.5$  &  $
 53^{+13}_{-15} $  & $ 69.5 \pm  5.6 $ & $  55.8 \pm 7.4  $ & $
 73.6 \pm 3.4 $ & $ 58.1 \pm  8.5 $ & $ 73.6 \pm 3.5 $ \\ &
 $n_s$ & $0.963 \pm 0.014$ & $0.955  \pm 0.014$ & $ 1.016 \pm 0.024 $ &
 $ 0.983 \pm  0.019 $ & $ 0.987 \pm  0.018 $ & $ 0.971 \pm  0.016 $ & $
 0.973 \pm 0.0149 $ \\ \hline Derived & & & & & & & & \\ &  $\Omega_{DE}(z = 0.0)$ & $0.73 \pm 0.03
 $ & $ < 0.77 \ (95 \% \ CL)$ & $ 0.72 \pm 0.04 $ & $ 0.63 \pm 0.06
 $ & $ 0.75 \pm 0.02 $ & $ 0.64 \pm 0.07 $ & $ 0.75 \pm 0.02 $ \\
 & $ t_0 $ [Gyr] & $13.8 \pm 0.1$ & $ 15.9^{+2.0}_{-1.7} $ & $ 13.5
 \pm 0.6 $ &  $ 16.4 \pm 1.2 $ & $ 13.5 \pm  0.5 $ & $ 15.9
 \pm 1.2 $ & $ 13.5 \pm 0.4 $ \\ \hline \hline
 \end{tabular}
 \end{center}
 \caption[Summary  of results  for  flat  $\Lambda$CDM  and 20  $w_i$
 parameters searches] {\footnotesize{ Summary
 of results. {\bf  WMAP7} refers to
 the  results  in  \cite{Larson11}   where only the WMAP 7-year  data  are
 used. O$\Lambda$CDM has $\Omega_k \neq 0$.  All other runs use 20 $w_i$ parameters with Prior I (II) imposing $-15 \leq w_i \leq 1$ ($-1 \leq w_i \leq 1$). The {\sc BASIC} data consists of WMAP7+SNIa+BBN while HST is the Hubble constant constraint $H_0 = 74.2 \pm 3.7 \Hunit$ and ACT is the CMB weak lensing spectrum measurement from ACT \cite{Das11}. Note the large range of ages in cases where the HST prior is not used and that $n_s = 1$ is compatible at $2\sigma$ with all dynamical dark energy runs.}}
 \label{cosmic_deg}
%\vspace{-0.15in}
 \end{table*}

It is a subtle combination of effects, wide coverage and multiple datasets, that gives good constraints on curvature in the case of general dark energy dynamics. One of the main reasons for this result is that while it is possible to match either the flat $\Lambda$CDM distances or expansion rate, $H(z)$, in a curved cosmos, it is not possible to do both simultaneously over an extended redshift range: $\Omega_k = 0$ is the only solution to the equation $\sin(\sqrt{-\Omega_k} \chi)/\sqrt{-\Omega_k} = \chi$, where $\chi = H_0\int dz'/H(z')$ is the usual flat $\Lambda$CDM comoving distance.  Hence, simultaneous low-redshift measurements of both Type Ia supernovae (SNIa) which constrain distances and the ISW effect, which constrains $H(z)$, provide constraints on curvature even in the absence of an $H_0$ prior. Our results are consistent with Fisher matrix projections \cite{Mortonson09}. %BB  
  
In addition to our \BASIC and \BASIC+ {\sc HST} runs, we also undertook runs with the CMB lensing results from the Atacama Cosmology Telescope (ACT) measurements \cite{Sherwin:2011gv} and the time-delay distance to the lens system B1608+656 at $z = 0.63$ \cite{Suyu10} as alternatives to the HST prior. However these two additional datasets did not significantly impact the constraint on $\Omega_k$. The impact of ACT CMB lensing is shown in the final two columns of Table (\ref{cosmic_deg}). 

We can also look at the effect of general dark energy dynamics and curvature on other parameters, such as the scalar spectral index, $n_s$, where assuming flat $\Lambda$CDM, there is evidence for $n_s < 1$ at a significance of about $3\sigma$ for WMAP \cite{Komatsu08} or over $5 \sigma$ for the latest SPT data \cite{SPT}. As shown in \fig(\ref{okns}) our results are all consistent with the Harrison-Zel'dovich value of $n_s = 1$, due to the large possible ISW effects from both curvature and dark energy dynamics \cite{psisw}, illustrating how cosmological constraints on the early universe are tightly coupled to the late universe. 

{\it Conclusions --} We have  established constraints on the curvature allowing for a general evolution of the dark energy equation of state which does not rely on artificial breaking of the curvature-dynamics degeneracy. The WMAP7 and Union2 supernova data are sufficient to constrain the curvature to \priorIInohst ,  if quintessence-like behaviour are assumed but if crossing of the phantom divide $w(z)=-1$ is allowed then we find \priorInohst , at $2\sigma$. As shown Fig (\ref{okns}), one may suspect that the strong constraining power of $H_0$ implies that the constraints on $\Omega_K$ are primarily coming from low $z$. In fact $H_0$ is important primarily through its control of the Hubble scale which sets the overall scale for $d_L(z)$. With the HST prior $H_0 = 74.2 \pm 3.7 \Hunit$ added we find \priorIhst , implying that the curvature-dynamics degeneracy does not change the results of the standard $\Lambda$CDM picture although we emphasise that our curvature constraints are no longer primarily driven by the distance to the surface of last scattering, as illustrated in Fig. (\ref{cls}), but rather the full shape of the power spectrum at both small and large angular scales.  

While these results do not assume that we know the dynamics of the dark energy, they do, however, assume that gravity is described by General Relativity and the speed of sound of the dark energy is unity. In this sense our results still leave a last loop-hole. 
Fortunately number counts, $N(z)$, time delays, lensing and growth of structure measurements depend on curvature very differently compared with distances and hence future data will allow us to close this loop-hole too \cite{Bernstein05, Mortonson09, Shafieloo2011, HCCB}. The key will be to extract accurate measurements of these observables that are truly independent both of the assumptions of flatness, $\Lambda$CDM, and General Relativity.

\paragraph*{Acknowledgements}
%\scriptsize
We thank Ren\'ee Hlozek for early work on the ideas in this paper and Pedro Ferreira, Dragan Huterer, Martin Kunz, Samuel Leach, Olga Mena, David Parkinson and Adam Riess for discussions. YF thanks AIMS for supporting a visit in which part of this work was carried out. PO and BB are funded by the SKA (SA). BB is funded by the National Research Foundation. We acknowledge use of the facilities of the Centre for High Performance Computing, Cape Town. 

%\vspace{-0.2in}

%\clearpage

%\clearpage

%\appendix

\end{document}